# COUPLING IMPEDANCES AND HEATING DUE TO SLOTS IN THE KEK B-FACTORY

Sergey S. Kurennoy* and Yong Ho Chin
KEK, 1-1 Oho, Tsukuba-shi, Ibaraki-ken, 305, Japan

**Abstract**

The longitudinal and transverse coupling impedances produced by the long slots in the Low Energy Ring of KEK B-factory are calculated. The power dissipated inside the vacuum chamber due to the fields scattered by the slots is evaluated using results for the real part of the coupling impedance. Estimates are made for the power flow through the slots to the pumping chamber.

*Permanent address: University of Maryland, College Park, MD 20742, USA

# 1 Introduction

The KEK B-factory is an electron-positron collider with unequal beam energies for study of B-meson physics, which is now in the process of design at the KEK [1]. The dominant issues in the KEKB from the point of view of beam instabilities are high beam currents (2.6 A in the Low Energy Ring (LER) and 1.1 A in the High Energy Ring (HER)) to achieve high luminosity of $10^{34}$ cm$^{-2}$s$^{-1}$, and a short bunch length ($\sigma_z = 4$ mm). Some relevant parameters of the machine are listed in Table 1.

Table 1: KEKB Machine Parameters

| Case | LER | HER | |
|---|---|---|---|
| Circumference | 3016.26 | | m |
| Energy | 3.5 | 8.0 | GeV |
| Particles | $e^+$ | $e^-$ | |
| Number of bunches | 5120 | | |
| Bunch spacing | 0.6 | | m |
| Betatron tune (h/v) | 45.52/45.08 | 46.52/46.08 | |

Because of the short bunch length, the beam can see coupling impedances up to very high frequencies, about 20 GHz, and due to the high currents it can create essential heat deposition via higher-order mode (HOM) losses. So, it is important to know the coupling impedances and loss factors of the chamber elements up to very high frequencies.

In the present paper, we analyze the coupling impedances of pumping and RF slots in the LER. The LER vacuum chamber is a circular pipe with inner radius $b = 50$ mm and wall thickness $t = 6$ mm. About 1800 m of the beam pipe have long pumping slots to connect it with the pumping chamber containing the NEG pumps. The slots are of rectangular shape with rounded ends, with width $w = 4$ mm and length $l = 100$ mm. They are located in groups of a dozen of parallel slots near the bottom of the beam pipe, about 20 slots per meter, with the total number of pumping slots in the LER about 36000.

Another type of slots in the LER vacuum chamber are narrow longitudinal slots (1.2 mm$\times$35 mm) in RF connectors. The number of slots per connector is 40, with about 100 connectors in the ring.

The paper is organized as follows. In Section 2, the low-frequency (below the chamber cutoff) impedances are calculated. Section 3 deals with high-frequency impedances and loss factors. The real part of the slot impedance is calculated, an expression for the loss factor of the long slot is derived, and power losses due to slots are evaluated. We estimate also the power flow through slots into the pumping chamber in this Section. Section 4 gives our conclusions.

# 2 Impedance at Low Frequencies

The imaginary part of the longitudinal and transverse coupling impedances of a slot is given by analytical formulae of Ref. [2]. The longitudinal impedance of the slot in the chamber with the circular cross section of radius $b$ is

$$Z(\omega) = -iZ_0 \frac{\omega}{c} \frac{(\alpha_{mag} + \alpha_{el})}{4\pi^2 b^2} \ , \tag{1}$$

where $Z_0 = 120\pi \ \Omega$, and $\alpha_{el}$, $\alpha_{mag}$ are the electric and magnetic polarizabilities of the slot, respectively. The transverse impedance is

$$\vec{Z}_{\perp}(\omega) = -iZ_0 \frac{\alpha_{mag} + \alpha_{el}}{\pi^2 b^4} \vec{a}_h \cos(\varphi_h - \varphi_b) \ , \tag{2}$$

where $\vec{a}_h$ is the unit vector directed to the slot in the transverse cross section of the chamber, $\varphi_h$ and $\varphi_b$ are azimuthal angles of the hole and beam in this cross section. The sum of polarizabilities for a long longitudinal slot of length $l$ and width $w$ ($w \ll l$), with rounded ends in a thin wall is [3]

$$\alpha_{mag} + \alpha_{el} = w^3(0.1334 - 0.0500w/l) \ . \tag{3}$$

For a thick wall, Eq. (3) should be multiplied by a thickness correction factor [4]. This factor is about 0.6 for long elliptic holes [4], but it is unknown for the case of a very long slot with rounded ends and parallel sides. For a cautious estimate one can take this factor equal to the ratio of the leading terms of the polarizabilities for an infinitely long slot in a thick and thin wall, namely $8/\pi^2$ [3].

Due to additivity of the impedances at frequencies below the chamber cutoff frequency (2.3 GHz), the analytical results above give reliable estimates

Table 2: Reactive Impedances of LER Slots

| Slots | $\lvert Z/n \rvert / \ \Omega$ | $\lvert Z_\perp \rvert / \ (\mathrm{k}\Omega/\mathrm{m})$ |
|---|---|---|
| Pumping | 0.002 | 1.5 |
| RF connectors | $10^{-4}$ | 0.035 |

of the LER total coupling impedances due to slots in this frequency range, see Table 2.

Since the pumping slots in LER are located near the bottom of the chamber, they contribute mostly to the vertical transverse impedance (the value in Table 2), cf. Eq. (2).

One should note also that the impedance behavior described by Eqs. (1)-(2) is in general valid up to even higher frequencies so far the wavelength is large compared to the slot width, but the wavelength can be smaller than the slot length. It has been demonstrated in Ref. [5]. An additional feature of the slot impedance in this frequency range (above the cutoff) is the presence of resonance peaks near the cutoff frequencies of all eigenmodes of the beam pipe, as it will be discussed in the next section.

# 3 Impedance at High Frequencies

## 3.1 The real part of the slot coupling impedance

The real part of the hole impedance in the frequency range where the wavelength is large compared to the hole (slot) dimensions, is proportional to the sum of polarizabilities squared, see Ref. [2]:

$$Re\, Z(\omega) = Z_0 \left(\frac{\omega}{c}\right)^4 \frac{(\alpha_{mag}^2 + \alpha_{el}^2)}{24\pi^3 b^2} \ . \tag{4}$$

For the frequency range where the wavelength can be smaller than the slot length but is still larger than its width, the analytical expression for the real part of the longitudinal coupling impedance for the long slot in the wall of a circular beam chamber has been obtained [5]:

$$Re\, Z(\omega) \quad = \quad \frac{2 Z_0 \omega (\alpha_{mag}/l)^2}{\pi b^4} \times$$

$$
\begin{aligned}
\times & \left\{ \sum_{n=0}^{\infty} \sum_{m=1}^{\infty} \frac{1}{(1+\delta_{n,0})\sqrt{\omega^2-\omega_{nm}^2}} \theta\left(\frac{\omega}{\omega_{nm}}-1\right) \cdot \right. \\
& \cdot \left[ \sin^2 \frac{l}{2c}(\omega - \sqrt{\omega^2-\omega_{nm}^2}) + \sin^2 \frac{l}{2c}(\omega + \sqrt{\omega^2-\omega_{nm}^2}) \right] \\
+ & \sum_{n=1}^{\infty} \sum_{m=1}^{\infty} \frac{n^2}{(\mu_{nm}'^2 - n^2)\sqrt{\omega^2-\omega_{nm}'^2}} \theta\left(\frac{\omega}{\omega_{nm}'}-1\right) \cdot \\
& \left. \cdot \left[ \sin^2 \frac{l}{2c}(\omega - \sqrt{\omega^2-\omega_{nm}'^2}) + \sin^2 \frac{l}{2c}(\omega + \sqrt{\omega^2-\omega_{nm}'^2}) \right] \right\} ,
\end{aligned} \tag{5}
$$

where $\delta_{n,0}$ is the Kronecker symbol, $\theta(x)$ is the Heaviside function, $\omega_{nm} = \mu_{nm} c/b$ and $\omega'_{nm} = \mu'_{nm} c/b$ are the cutoff frequencies of respectively TM- and TE-eigenmodes of the circular waveguide. Here $\mu_{nm}$ is the $m$th root of the Bessel function $J_n(x)$, and $\mu'_{nm}$ is the same for its derivative $J'_n(x)$. Equation (5) has been derived for a long slot in a perfectly conducting wall, and it is taken into account that for a long slot $\alpha_{mag} \simeq -\alpha_{el} \propto w^2 l$, see e.g. in Ref. [3]. The resonance peaks of the impedance near the cutoffs will be, of course, the finite ones if the wall conductivity is finite, and the radiation through slots is taken into account. For lower frequencies, the summation in Eq. (5) can be carried out analytically [5], and it is reduced to Eq. (4).

The behavior of the slot impedance as a function of frequency according to Eq. (5) is shown in Fig. 1 for the parameters of KEKB LER pumping slots. Figure 2 shows the frequency region near the chamber cutoff in more detail, and compares the real part with the imaginary one given by Eq. (1) (i.e., without contributions from peaks of the real part).

However, we are mostly interested in the power loss due to slots and in the power flow through the slots to the pumping chamber.

## 3.2 Power losses

The power per unit length of the beam chamber dissipated due to beam fields scattered by the slots can be expressed as

$$
P' = \frac{N_{sl}}{S_b} f_{rev} q_b^2 k \ , \tag{6}
$$

where $N_{sl} = 36000$ is the total number of slots in the LER, $S_b = 0.6$ m is the bunch spacing, $f_{rev} = c/(2\pi R)$ is the revolution frequency, $q_b$ is the bunch charge, and $k$ is the loss factor per slot. The loss factor can be obtained by integrating $Re\, Z$ with the bunch spectrum:

$$k = \frac{1}{\pi}\int_0^\infty d\omega Re\, Z(\omega)\exp\left[-\left(\frac{\omega\sigma}{c}\right)^2\right] \ , \tag{7}$$

where $\sigma = \sigma_z = 4$ mm is the r.m.s. bunch length. Substituting Eq. (5) into Eq. (7) yields

$$\begin{aligned} k = \frac{2Z_0 c(\alpha_{mag}/l)^2}{\pi^2 b^4 \sigma} \quad \cdot \quad & \left[\sum_{n=0}^\infty \sum_{m=1}^\infty \frac{1}{1+\delta_{n,0}} F\left(\frac{\mu_{nm}\sigma}{b}, \frac{l}{\sigma}\right) + \right. \\ & \left. + \sum_{n=1}^\infty \sum_{m=1}^\infty \frac{n^2}{\mu'^2_{nm} - n^2} F\left(\frac{\mu'_{nm}\sigma}{b}, \frac{l}{\sigma}\right)\ \right] \ , \end{aligned} \tag{8}$$

where

$$\begin{aligned} F(x,y) \ \equiv \ & \exp(-x^2)\int_0^\infty dz \exp(-z^2) \ \cdot \\ \cdot \quad & \left[\sin^2 \frac{y}{2}(\sqrt{z^2+x^2}-z) + \sin^2\frac{y}{2}(\sqrt{z^2+x^2}+z)\right] \ . \end{aligned} \tag{9}$$

It is easy to realize that for the values of arguments of function $F$ which are of interest in Eq. (8), namely $y = l/\sigma \gg 1$, one can replace fast oscillating functions in the square brackets of Eq. (9) by their average values. Then the integral is evaluated analytically:

$$F(x,y) \simeq \exp(-x^2)\frac{\sqrt{\pi}}{2} \ . \tag{10}$$

One can check numerically that there will be only small deviations within a few percent from this asymptotic for the lowest roots $\mu_{nm}, \mu'_{nm}$ entering Eq. (8), for the parameters of LER: $l/\sigma = 25$ and $b/\sigma = 12.5$, see in Fig. 3.

As a result, we get a simplified expression for the loss factor

$$\begin{aligned} k = \frac{Z_0 c(\alpha_{mag}/l)^2}{\pi^{3/2} b^4 \sigma} \quad \cdot \quad & \left\{ \sum_{n=0}^\infty \sum_{m=1}^\infty \frac{1}{1+\delta_{n,0}} \exp\left[-\left(\mu_{nm}\sigma/b\right)^2\right] + \right. \\ & \left. + \sum_{n=1}^\infty \sum_{m=1}^\infty \frac{n^2}{\mu'^2_{nm} - n^2} \exp\left[-\left(\mu'_{nm}\sigma/b\right)^2\right] \ \right\} \ . \end{aligned} \tag{11}$$

Let us introduce notation $S(b/\sigma)$ for the sum in the brackets in the last equation. We have calculated its value for the LER parameters, and the result is $S(12.5) = 34.7$. Since the wall thickness $t = 6$ mm is larger than the slot width $w = 4$ mm, one can use the polarizability for a long slot in a thick wall [3]: $\alpha_{mag}/l = w^2/(2\pi)$. Then Eq. (11) gives the loss factor per slot

$$k = \frac{Z_0 c w^4}{4\pi^{7/2} b^4 \sigma} S\left(\frac{b}{\sigma}\right) = 1.83 \cdot 10^{-4}\ \mathrm{V/pC}\ . \tag{12}$$

After substituting it to Eq. (7), we get the slot contribution to the power dissipated per unit length of the chamber

$$P' \simeq 30\ \mathrm{W/m}\ .$$

One can derive an asymptotic of the function S(x) defined above in the limit of $x \gg 1$. Using asymptotic expressions [6] of Bessel roots

$$\begin{aligned} \mu_{nm} &= (m + \frac{n}{2} - \frac{1}{4})\pi + O(\frac{n^2}{m})\ , \\ \mu'_{nm} &= (m + \frac{n}{2} - \frac{3}{4})\pi + O(\frac{n^2}{m})\ , \end{aligned} \tag{13}$$

valid for $m \gg n$, and replacing the sums by integrals, we get an asymptotic

$$S(x) \simeq \frac{x^2}{\pi^2} + \frac{x^2}{16\pi}\ , \tag{14}$$

where the first term corresponds to the contribution from TM-modes (the first sum in Eq. (11)), and the second one to that from TE-modes (the second sum in (11)). For the LER parameters ($x = b/\sigma = 12.5$), Eq. (14) gives $15.8 + 3.1 = 18.9$ instead of the correct value $34.7 = 16.7 + 18.0$, that means asymptotic (14) is not good for the TE part. The reason is that due to the weight factor $n^2/(\mu'^2_{nm} - n^2)$ main contributions to the TE sum come from the lowest roots, $m \ll n$, and for these roots Eqs. (13) do not work, of course. One can see it from a numerical summation in Eq. (11) for the LER parameters, in which the lowest roots, $m \ll n$, with $n$ up to 20, contribute most of the result, and maximum contributions come from $n$ around 7. One can estimate this additional contribution using a rough approximation for the lowest roots, $\mu'_{nm} \simeq n + \pi m - 3/2$. Restricting sums over $m$ by values up to $n$, and replacing sums by integrals, we get an additional term of asymptotic

$$\Delta S(x) \simeq \frac{x^{5/2} \log x}{144}\ , \tag{15}$$

that recovers extra 9.7 for the LER parameters. A more elaborated study could probably give a better approximation, but, nevertheless, Eqs. (14)-(15) give an idea of scaling of the loss factor for large values of $x = b/\sigma$. It should be noted, however, that the whole consideration above works only for the case when the bunch length is not small compared to the slot width, $\sigma \geq w$, since Eq. (5) is applicable only in this frequency range.

One can see that the loss factor for a long slot is independent of the slot length as far as this length is large compared to the bunch length. So, to get the result for RF connector slots, we just substitute the different slot width, $w = 1.2$ mm, into Eq. (12). Multiplying it by the number of slots $N_1 = 40$ in one connector, we get the loss factor per connector

$$k = 6 \cdot 10^{-5} \text{ V/pC} .$$

## 3.3 Power flow through slots to the pumping chamber

### 3.3.1 Power flow due to direct beam fields on the slots

When the beam field illuminates a slot, it produces scattered fields both inside the beam pipe and outside, in the pumping chamber. The energy radiated by these outside fields into the pumping chamber can be calculated in the same way as for the beam pipe. One should only use so-called "external" polarizabilities of slots instead of the "internal" ones which we used above, cf. [4]. For the case of the thick wall (thickness $t$ is larger than the slot width $w$) these "external" polarizabilities are much smaller than the "internal" ones:

$$\alpha^{ext}_{el,mag} \simeq \exp\left(-\pi t/w\right) \, \alpha_{el,mag} \; . \tag{16}$$

Since the power loss is proportional to the polarizabilities squared, one can easily realize that the power flow to the pumping chamber due to this source is negligible:

$$P'_1 \simeq \exp\left(-2\pi t/w\right) \, P'_{in} \simeq 3 \text{ mW/m} .$$

### 3.3.2 Power flow from transverse currents due to the beam betatron motion

One should note that the relation (16) works only for the electric and transverse magnetic polarizabilities. The term "transverse" here means that the beam magnetic field is directed perpendicular to the slot largest dimension, and this is correct for the case of the beam longitudinal motion when there is only the azimuthal component of the beam magnetic field. However, if the transverse motion of the beam is taken into account there is also some longitudinal magnetic field of the beam on the slot. In this case the longitudinal magnetic dipole moment is induced on the slot. It is proportional to the longitudinal magnetic polarizability of the slot, which is equal to [3]

$$\alpha_z = \frac{\pi l^3}{24} \left( \ln \frac{8l}{w} + \frac{\pi t}{2w} - \frac{7}{3} \right)^{-1} , \tag{17}$$

and is much larger than the transverse one, $\alpha_\varphi = w^2 l/(2\pi) + O(w^3)$, for long slots. In addition, its "internal" and "external" values are almost equal, because the magnetic field parallel to the slot easily penetrate through it. So, we have to estimate the power flow due to this effect.

Since the power radiated in TE-modes $P'_H \propto Re\, Z \propto M^2$, where $M = \alpha_{mag} H$ is the effective magnetic dipole moment induced by the beam magnetic field $H$ on the slot, we will consider the ratio

$$r = \frac{M_z}{M_\varphi} = \frac{\alpha_z H_z}{\alpha_\varphi H_\varphi} . \tag{18}$$

The ratio of fields $H_z/H_\varphi$ can be approximated by the ratio of the beam current components, $j_\perp/j \simeq \nu/n$, where $\nu \simeq 45$ is the betatron frequency, and $n = \omega/\omega_0$ is the longitudinal harmonic number, with $\omega_0 = c/R$ being the revolution frequency. Then

$$r = \frac{\pi^2 l^2}{12 w^2} \left( \ln \frac{8l}{w} + \frac{\pi t}{2w} - \frac{7}{3} \right)^{-1} \frac{\nu \omega_0}{\omega} , \tag{19}$$

which is about 0.2 at the beam pipe cutoff frequency. We are interested in frequencies above the chamber cutoff, and because (19) decreases with frequency increase, one can conclude that the power flow into the pumping chamber $P'_2$ from this source is less than 1/25 of the power radiated in TE-modes into the beam pipe, which is about one half of the total $P'$, i.e. about 15 W/m (the second sum in Eq. (11)). So,

$$P'_2 < 0.6 \text{ W/m} .$$

### 3.3.3 Power flow due to the fields scattered by slots to the beam pipe

It is mentioned above that the total power $P'$ radiated by slots into the beam pipe is divided almost evenly between TM- and TE-modes: $P'_E \simeq P'_H \simeq P'/2 = 15$ W/m. These modes propagate in the beam pipe and reach following slots. For TM-waves penetration through the longitudinal slots is exponentially small, see above. However, TE-waves have a longitudinal magnetic field (the wall currents are transverse to the slots), and contribute to the energy flow into the pumping chamber. The minimal (optimistic) estimate for that power flow would be

$$P_3'^{min} = \frac{A_{sl}}{A_{wall}} P'_H \simeq 0.4 \text{ W/m} ,$$

where $A_{sl}/A_{wall} = 0.025$ is the fraction of the wall surface occupied by slots. On the other hand, the most pessimistic (maximal) estimate is to assume that the total energy of TE-modes will leak out through slots:

$$P_3'^{max} = P'_H \simeq 15 \text{ W/m} .$$

Of course, a part of the energy in TE-modes will be lost in the walls of the beam pipe, so the last number gives the upper estimate.

# 4 Conclusions

The calculated low-frequency coupling impedances of the pumping slots in the Low Energy Ring of the KEKB contribute a few percent to the present low-frequency impedance budget.

The power losses in the LER beam pipe due to the pumping slots are not negligible and exceed that due to the resistive wall. It depends strongly on the slot width and almost independent of its length, as far as the last one is much larger than the bunch length. The analytical expression for the loss factor of a long slot for short bunches is derived.

The estimates of the power flow through the slots into the pumping chamber show that the most important contribution is that due to TE-modes scattered by the slots to the beam pipe. The upper estimate for this contribution is 15 W/m. A more accurate result should be obtained by a further study. This study would also help to take into account the power leakage to the pumping chamber from waves generated by other discontinuities of the vacuum chamber.

**Acknowledgements**
One of the authors (S.K.) would like to thank the KEK B-factory team, and especially Prof. Shin-Ichi Kurokawa and Prof. Yong Ho Chin, for their hospitality during his stay at KEK.

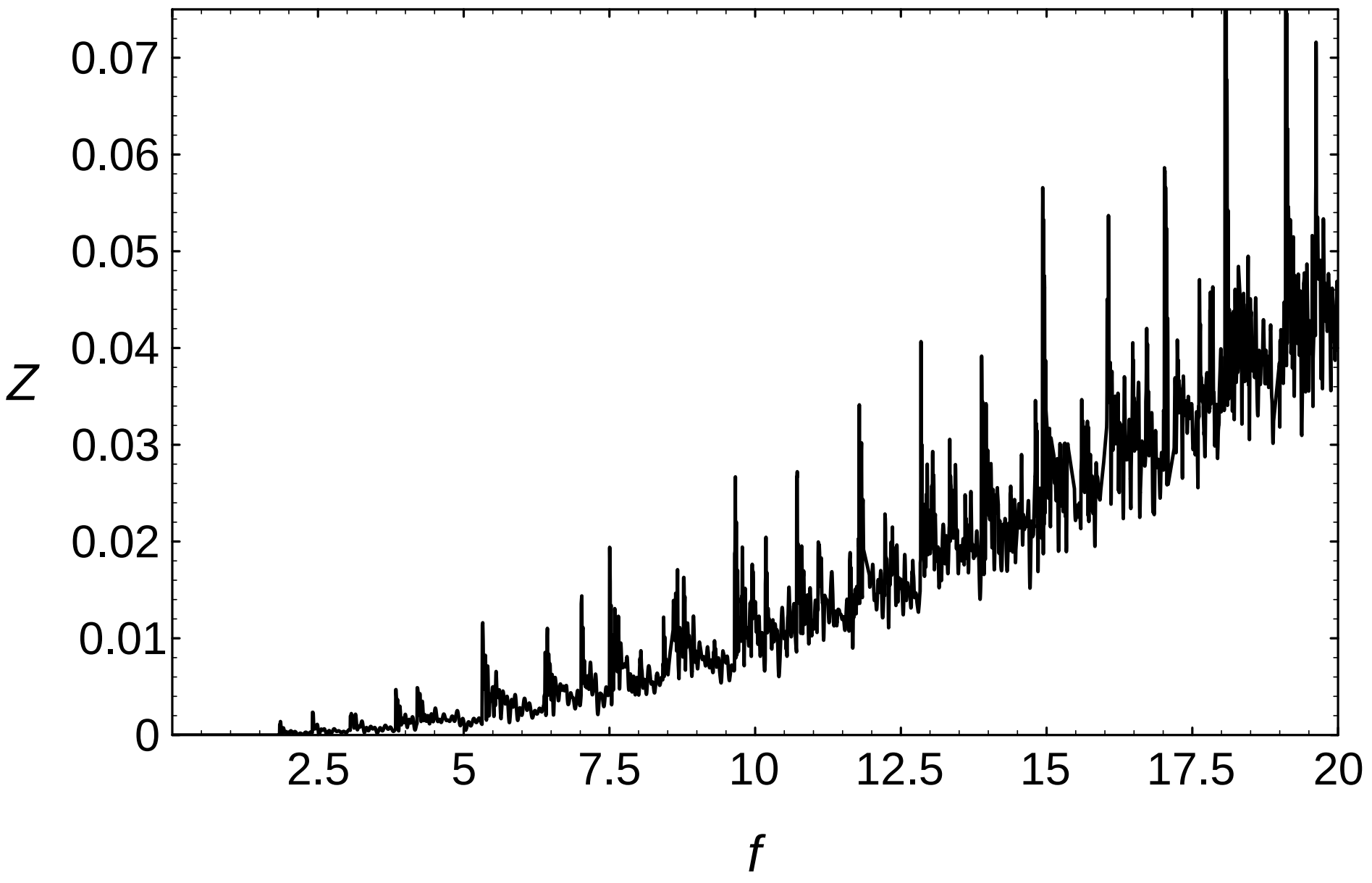

Figure 1: The real part of the slot longitudinal impedance ($Z/\Omega$) versus frequency ($f$/GHz).

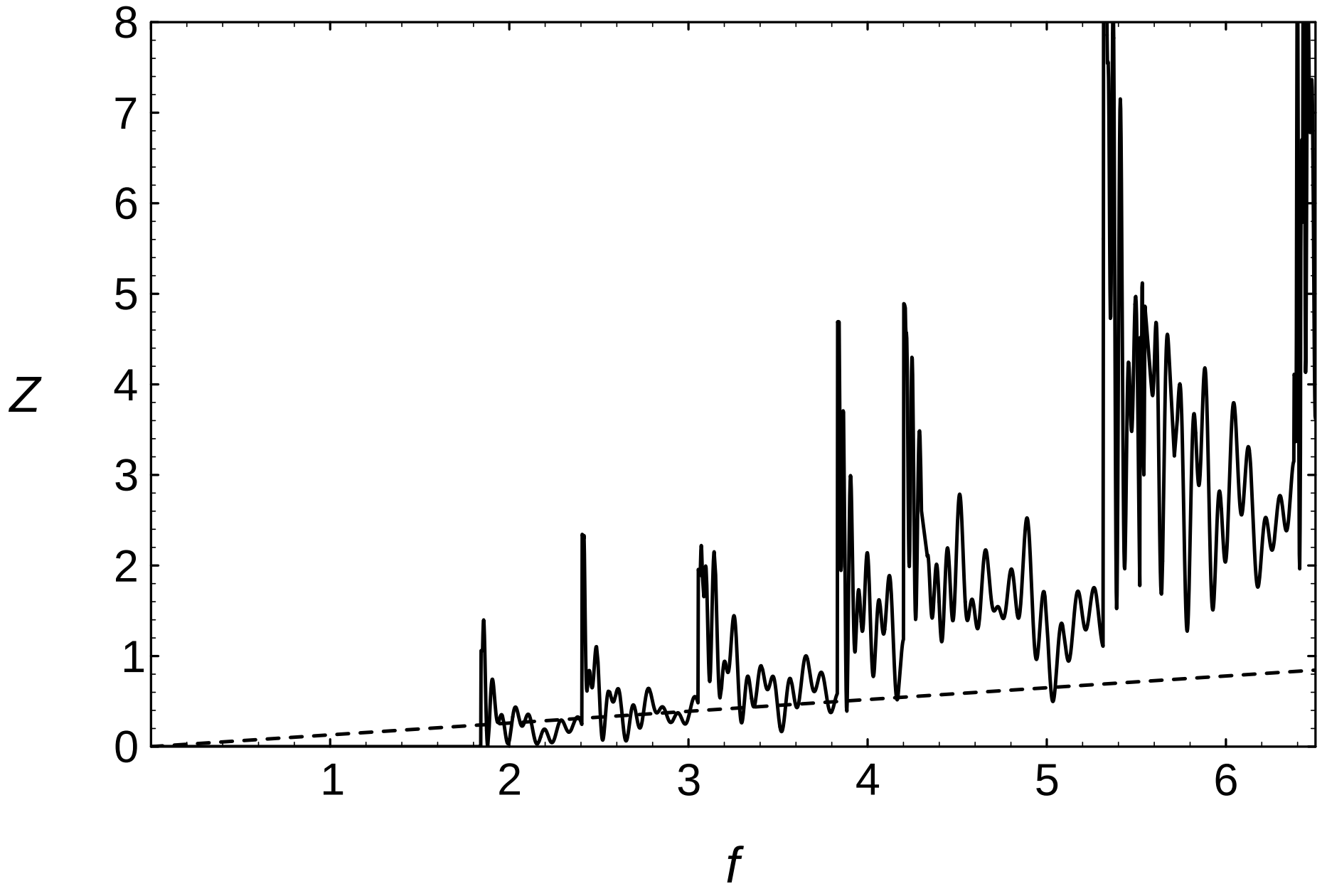

Figure 2: The longitudinal impedance of the slot ($Z/\mathrm{m}\Omega$) versus frequency ($f/\mathrm{GHz}$) near cutoff: the real part (solid line) and the imaginary part (dashed).

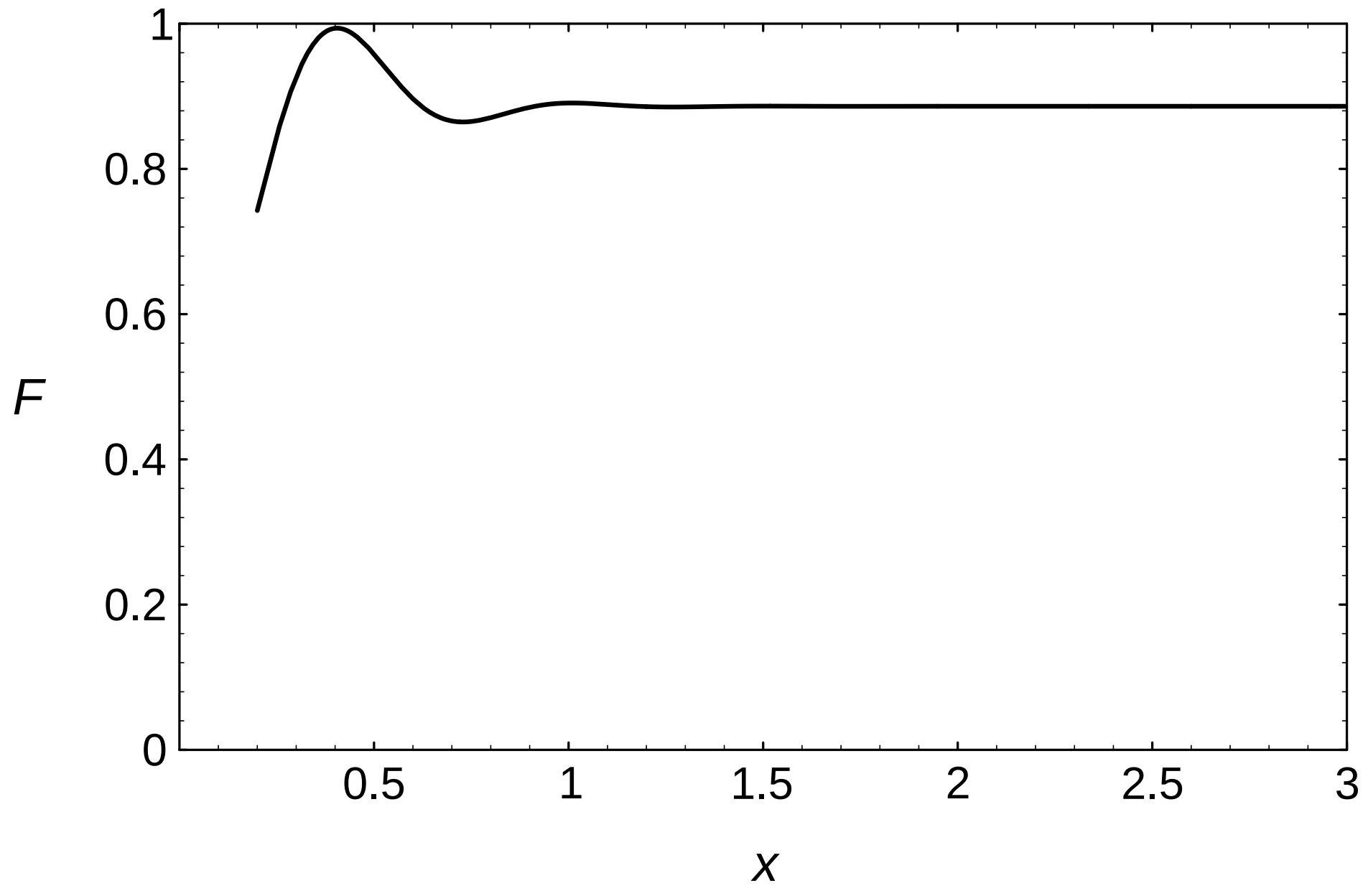

Figure 3: Function $F(x, y)$ versus $x$ for $y = 25$.